\documentclass[conference]{IEEEtran}
\IEEEoverridecommandlockouts

\usepackage{cite}
\usepackage{amsmath,amssymb,amsfonts}
\usepackage{algorithmic}
\usepackage[ruled, linesnumbered]{algorithm2e}
\usepackage{graphicx}
\usepackage{textcomp}
\def\BibTeX{{\rm B\kern-.05em{\sc i\kern-.025em b}\kern-.08em
    T\kern-.1667em\lower.7ex\hbox{E}\kern-.125emX}}

\usepackage[normalem]{ulem}

\usepackage{url}

\usepackage{subfig}
\usepackage{multirow}
\usepackage{lipsum}
\usepackage{wrapfig}

\usepackage{makecell}
\usepackage{xspace}
\usepackage{pifont}

\usepackage{caption}
\usepackage{booktabs}

\DeclareMathAlphabet\mathbfcal{OMS}{cmsy}{b}{n}

\usepackage{color}
\usepackage[dvipsnames,svgnames,table]{xcolor}

\usepackage{lipsum}

\newcommand{\sym}[1]{\textsf{#1}}

\newcommand{\name}{\sym{Catwalk}\xspace}
\usepackage{adjustbox}
\usepackage{bm}
\usepackage{soul}

\newcommand{\thetitle}{Catwalk: Unary Top-K for Efficient Ramp-No-Leak Neuron Design for Temporal Neural Networks}

\begin{document}
\title{\thetitle}

 \author{\IEEEauthorblockN{Devon Lister}
 \IEEEauthorblockA{\textit{University of Central Florida} \\
 Orlando, FL, USA \\
 devon.lister@ucf.edu}
 \and
 \IEEEauthorblockN{Prabhu Vellaisamy}
 \IEEEauthorblockA{\textit{Carnegie Mellon University} \\
 Pittsburgh, PA, CMU \\
 pvellais@andrew.cmu.edu}
 \and
 \IEEEauthorblockN{John Paul Shen}
 \IEEEauthorblockA{\textit{Carnegie Mellon University} \\
 Pittsburgh, PA, CMU \\
 jpshen@andrew.cmu.edu}
 \and
 \IEEEauthorblockN{Di Wu}
 \IEEEauthorblockA{\textit{University of Central Florida} \\
 Orlando, FL, USA \\
 di.wu@ucf.edu}
 }


\maketitle

\begin{abstract}
Temporal neural networks (TNNs) are neuromorphic neural networks that utilize bit-serial temporal coding.
TNNs are composed of columns, which in turn employ neurons as their building blocks.
Each neuron processes volleys of input spikes, modulated by associated synaptic weights, on its dendritic inputs. 
Recently proposed neuron implementation in CMOS employs a Spike Response Model (SRM) with a ramp-no-leak (RNL) response function and assumes all the inputs can carry spikes. 
However, in actual spike volleys, only a small subset of the dendritic inputs actually carry spikes in each compute cycle. This form of sparsity can be exploited to achieve better hardware efficiency. 
In this paper, we propose a \name neuron implementation by relocating spikes in a spike volley as a sorted subset cluster via unary top-k.
Such relocation can significantly reduce the cost of the subsequent parallel counter (PC) for accumulating the response functions from the spiking inputs. This can lead to improvements on area and power efficiency in RNL neuron implementation. 
Place-and-route results show \name is $1.39\times$ and $1.86\times$ better in area and power, respectively, as compared to existing SRM0-RNL neurons.
\end{abstract}


\section{Introduction}
\label{sec:Introduction}

Neuromorphic computing, particularly represented by spiking neural networks (SNNs) \cite{tavanaei2019deep}, has emerged as an attractive alternative to conventional compute-intensive deep neural networks (DNNs) due to its brain-inspired approach for computational efficiency. Unlike DNNs, which transmit continuous-valued signals, neurons in SNNs communicate information through discrete spike or pulse events. Central to these architectures is the neuron model, which defines how neurons generate and respond to spikes.
Among various neuron models, the spike response model (SRM) has been widely utilized, especially in its simplified variant, the SRM0 model, characterized by a fixed spike generation threshold. The SRM0 neuron can employ several types of response functions, such as biexponential~\cite{gutig2006tempotron}, piecewise linear~\cite{maass1997networks}, step-no-leak~\cite{tnn_clustering, dong2018unsupervised, kheradpisheh2018stdp}, and, notably, the ramp-no-leak (RNL) function~\cite{tnn_online_learning, tnn_clustering}. The RNL function has recently gained significant interest due to its practical advantages in temporal neural network (TNN) applications \cite{tnn_microarch, tnn_online_learning, tnn_clustering, chaudhari2021unsupervised, vellaisamy2024tnngen, smith2022implementing, nair2022tnn7}. TNNs, a special class of SNNs, are capable of continuous \cite{tnn_online_learning} online learning and unsupervised clustering \cite{chaudhari2021unsupervised, tnn_clustering} by employing precise spike timing and biologically-plausible spike-timing dependent plasticity (STDP) local learning rule. This contrasts with the compute intensive global backpropagation methods predominantly used in traditional DNNs.

Existing TNNs adopt an SRM0 neuron model with an RNL response function~\cite{tnn_online_learning, tnn_clustering}. For brevity, we henceforth call this neuron model an \textit{SRM0-RNL neuron} in this paper. However, current SRM0-RNL neuron implementations \cite{tnn_microarch, vellaisamy2024tnngen, nair2022tnn7} employ worst-case scenarios for temporal signal processing and provides increased hardware resources for maximal spike density scenarios, which rarely occurs due to the inherent sparsity, i.e., just 0.1\%-10\% of total count of neurons are observed to fire biologically~\cite{silent_neuron, temporal_sparsity, measuring_sparsity}.

\begin{figure}[!t]
    \centering
    \includegraphics[width=\columnwidth]{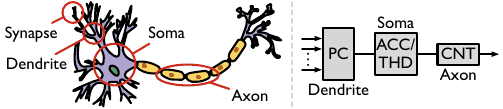}\\
    \caption{A bilogical neuron and and its circuit representation using ramp-no-leak response function. 
    PC (parallel counter) integrates all current incoming spikes modulated by the synaptic weights.
    ACC/THD is the soma (neuron body) that accumulates (ACC) all incoming potentials from dendrites and checks whether the accumulated potential surpasses a predefined threshold (THD).
    CNT (counter) is the axon output that fires a spike if the accumulated potential is higher than the threshold.
    The synapses are not shown here for simplicity.
    }
    \label{fig:neuron_mapping}
\end{figure}

To address these inefficiencies, this paper introduces \textit{\name}, a novel neuron architecture employing unary top-k to optimize spike aggregation at neuron inputs. \name leverages the inherent sparsity in temporal-coded spikes to cluster active spikes efficiently, enabling the replacement of conventional, fully provisioned parallel counter (PC), for accumulation of input response functions, with significantly more compact and efficient counterparts. Specifically, our contributions are:

\begin{itemize}
    \item We propose \name, a novel technique that relocates the top-k temporal spikes to optimize spike aggregation in neuron inputs for better hardware efficiency.
    \item A detailed evaluation against baseline designs, including SRM0-RNL neurons that employ conventional PCs and those that integrate unary sorting, demonstrates the clear advantages of our proposed \name neuron.
    \item Place-and-route evaluation at 45nm CMOS shows area and power improvements of $1.39\times$ and $1.86\times$, respectively, relative to existing SRM0-RNL neurons.
    
\end{itemize}




This paper is organized as follows.
Section~\ref{sec:Background} and Section~\ref{sec:Theory} review the background and motivate this work.
Then Section~\ref{sec:Architecture} describes the proposed neuron design.
The following Section~\ref{sec:Implementation} and Section~\ref{sec:Evaluation} evaluate the implementation.
Finally, Section~\ref{sec:Conclusion} concludes this work.

\section{Background}
\label{sec:Background}

\subsection{SRM0-RNL Neuron}
\begin{figure}[!t]
    \centering
    \subfloat[An existing SRM0-RNL neuron model~\cite{tnn_online_learning, tnn_clustering}.
    The input spikes $x_{\{0,1,2\}}$ are temporal-coded (red pulses), where the spike timing encodes the value.
    Note it is allowed that an input has no spike at all, representing a value of $\infty$, e.g., $x_{\{3\}}$.
    When a synapse receives an input spike, it will instantly trigger the RNL response function, which is described in Equation~\ref{eq:rnl_response_func}.
    This function generates a pulse whose width is identical to the weight value, indicated by the number of dots in blue.
    The soma will accumulate all these weight pulses as the membrane potential.
    Once the accumulated potential exceeds a threshold, the axon will fire an output spike, and the neuron will be reset.
    ]{\label{fig:neuron_model_tnn}\includegraphics[width=\columnwidth]{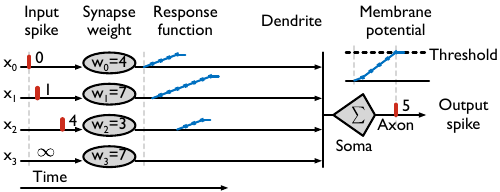}}\\
    \subfloat[Our proposed \name neuron model with unary top-k. 
    \name inserts unary top-k at the dendrite, which relocates and clusters the spikes together.
    Our spike relocation reduces the number of valid spike volleys without alternating the number of valid spikes, reducing the cost of PC in the dendrite.
    ]{\label{fig:neuron_model_catwalk}
       	\includegraphics[width=\columnwidth]{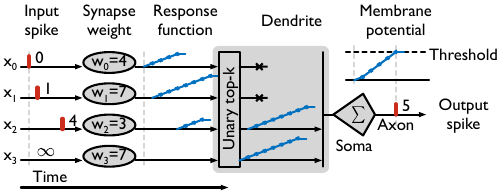}}\\
    \caption{Comparison of an existing SRM0-RNL neuron model and our proposed \name neuron model for TNNs.
    }
    \label{fig:neuron_model}
\end{figure}

SRM0-RNL neurons closely follow the formulation of biological neurons (Fig.~\ref{fig:neuron_mapping}), and existing TNNs encode information through precise spike timings rather than spike rates~\cite{tnn_online_learning, tnn_clustering}. 
The mechanism of this neuron model is explained further in Fig.~\ref{fig:neuron_model_tnn} from prior works~\cite{tnn_online_learning, tnn_clustering}, and is contrasted with our \name neuron model, depicted in Fig.~\ref{fig:neuron_model_catwalk}. 
Each post-synaptic neuron is fed by incoming temporal-coded spikes from all preceding neurons (pre-synaptic neurons), where the synapse responses are accumulated. The axon fires a spike if the accumulated membrane potential exceeds a certain threshold. 
In the existing SRM0-RNL implementations, all incoming responses are summed during potential accumulation~\cite{tnn_microarch}.
The RNL response function is provided in Equation~\ref{eq:rnl_response_func}, where $w$ is the weight value and $t$ refers to time.
Therefore, the RNL response function essentially creates a pulse of width $w$ over time, to be accumulated later.
\begin{equation}
\label{eq:rnl_response_func}
\begin{aligned}
\rho(w, t) & =0 & & \text { if } t<0 \\
& =t+1 & & \text { if } 0 \leq t<w \\
& =w & & \text { if } t \geq w
\end{aligned}
\end{equation}

\subsection{Unary Sorting}

\begin{figure}[!t]
    \centering
    \subfloat[Min and max via temporal coding.]{\label{fig:min_max_tc}
       	\includegraphics[width=\columnwidth]{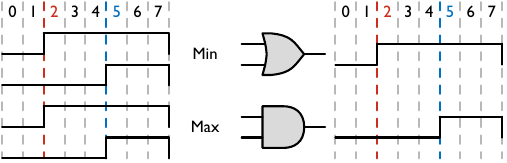}}\\
    \subfloat[Compare-and-swap unit via min and max. 
    This is also a 2-input bitonic sorter.
    Sorting more inputs can be done by repeating this unit recursively.]{\label{fig:bitonic_sorting}
       	\includegraphics[width=\columnwidth]{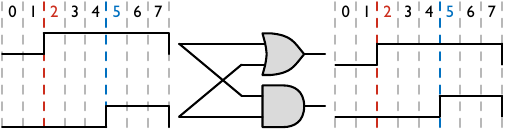}}\\
    \caption{Unary sorting via temporal coding.
    This example encodes unary data in a leading-0 mode. 
    The timing of the rising edge marks the data value, as colored by red and blue.
    }
    \label{fig:unary_sorting_tc}
\end{figure}

Unary data representations, either rate-coded or temporal-coded, have been used to perform various operations, including multiplication, addition~\cite{ugemm_paper}, division, square root~\cite{2019dac_instream}, min, max~\cite{Tzimpragos-ASPLOS19}.
Among these, bitonic sorting on temporal-coded unary data (e.g. temporal spikes in SRM0-RNL neurons) can be implemented via simple AND and OR gates~\cite{unary_sorting_paper}. 
An example of unary sorting is shown in Fig.~\ref{fig:unary_sorting_tc}.
Fig.~\ref{fig:min_max_tc} shows the use of simple AND and OR gates to perform min and max operations with temporal coding, respectively. 
Fig.~\ref{fig:bitonic_sorting} illustrates the basic compare-and-swap unit for unary sorting. 
The input temporal signals are ranked in order at the output, with the larger values clustered at the bottom. 
The simplicity of unary compare-and-swap units offers opportunities in integration with SRM0-RNL neurons, leading to a more optimized design.

\section{Motivation and Opportunity}
\label{sec:Theory}

The existing SRM0-RNL neuron design assumes the worst-case scenarios.
For an $n$-input neuron, the PC must accumulate all $n$ inputs, even if temporal spikes are not present for some inputs.
This worst-case neuron design does not consider the biological aspect of neurons, i.e. it is de facto that neuron spikes are extremely sparse and only $0.1\%\sim10\%$ of total neurons are actively spiking in any given compute cycle ~\cite{silent_neuron, temporal_sparsity, measuring_sparsity}.
Such a worst-case neuron design overprovisions hardware resources, incurring suboptimal efficiency.

Since the inputs of SRM0-RNL neurons are temporal coded, there exist opportunities to find the inputs with effective spikes using the unary logic in Fig.~\ref{fig:unary_sorting_tc}, and then use a more lightweight PC to aggregate these effective spikes.

In this work, we propose \name to take advantage of this existing technique and optimize spike aggregation for more efficient SRM0-RNL neurons in TNNs.
Our proposed \name neuron model ingests all temporal spikes from pre-synaptic neurons and relocates them, so that all active spikes are clustered together.
Then we can substitute the original full PC with a more lightweight version. 
As long as the cost of the spike relocation and the new PC is less than the original full PC, \name offers hardware efficiency gains.
Given the high neuronal sparsity within actual workloads, \name should not cause significant accuracy concerns. More experimental work is needed to validate this.

\section{\name Neuron via Unary Top-K}
\label{sec:Architecture}

In this section, we describe the hardware design of the proposed \name neuron using unary top-k.

\subsection{\name Neuron Microarchitecture}

\begin{figure}[!t]
    \centering

    \subfloat[Microarchitecture of an existing SRM0-RNL neuron.
    This example uses a 16-input PC as the dendrite to accumulate all possible input spikes (reproduced from~\cite{tnn_microarch}).
    The spikes are accumulated to the membrane potential register in the soma, which is then compared to a threshold to determine whether an output spike shall fire.
    The counter in the axon produces an 8-cycle pulse if an output spike occurs.]{\label{fig:neuron_uarch_tnn}\includegraphics[width=\columnwidth]{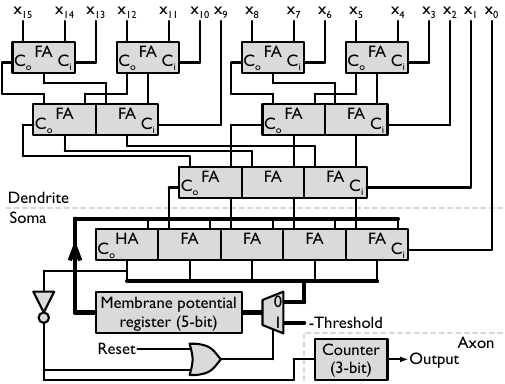}}\\
    \subfloat[Microarchitecture of our proposed \name neuron.
    This example feeds on 16 inputs and selects top-2 outputs.
    Instead of a large PC, the dendrite is now implemented using unary top-k, based on the compare-and-swap units (Fig.~\ref{fig:bitonic_sorting}) and half compare-and-swap units.
    The half compare-and-swap units does not have the dash gate, whose output is no longer needed.
    The design of soma and axon remains identical to that in existing SRM0-RNL neurons (Fig.~\ref{fig:neuron_uarch_tnn}).]{\label{fig:neuron_uarch_catwalk}
       	\includegraphics[width=\columnwidth]{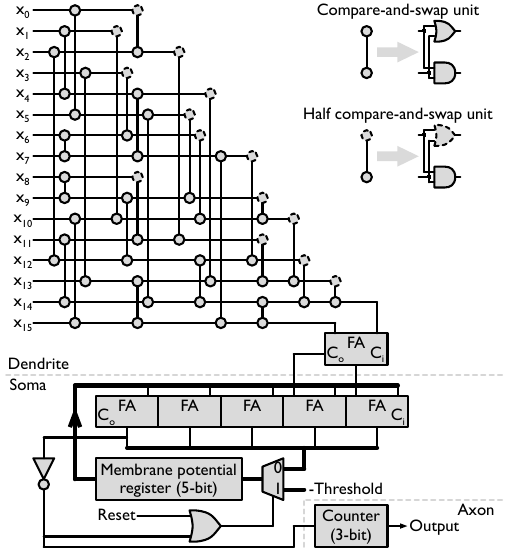}}\\
    \caption{Microarchitecture comparison of existing SRM0-RNL neuron and our proposed \name neuron.
    }
    \label{fig:neuron_uarch}
\end{figure}

We show the microarchitecture of the existing SRM0-RNL neuron and our \name neuron with unary top k in Fig.~\ref{fig:neuron_uarch}.
As illustrated in Fig.~\ref{fig:neuron_model_catwalk}, the key idea of \name is to identify the valid temporal spikes, which simply means finding the sparse bit ones at each clock cycle from an input collection of bit ones and zeros.
We coin this function as \textit{unary top k}.
As shown in Fig.~\ref{fig:neuron_uarch_tnn}, existing SRM0-RNL neurons employ large PCs for accumulating response functions of input dendrites, requiring $n-1$ full adders for $n$ inputs~\cite{tnn_microarch}.
Our \name neuron simply replace the large PC in the dendrite with unary top-k and smaller PC, with no other changes to the soma and axon.
As TNNs integrate multiple SRM0-RNL neurons into one TNN column~\cite{tnn_online_learning, tnn_clustering, tnn_microarch}, \name is a plug-and-play component that contributes to overall improvements in TNN efficiency.

\subsection{Unary Top-K}

\begin{figure}[!t]
    \centering
    \subfloat[Top-2 bitonic (24/19/6).]{\label{fig:top_2_8_bitonic}
       	\includegraphics[width=0.46\columnwidth]{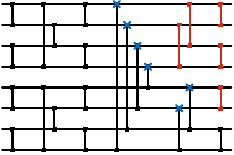}}
    \hspace{10pt}
    \subfloat[Top-4 bitonic (24/20/4).]{\label{fig:top_4_8_bitonic}
       	\includegraphics[width=0.46\columnwidth]{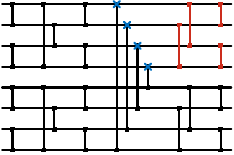}}\\
    \subfloat[Top-2 optimal (19/14/6).]{\label{fig:top_2_8_optimal}
       	\includegraphics[width=0.46\columnwidth]{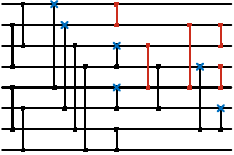}}
    \hspace{10pt}
    \subfloat[Top-4 optimal (19/18/4).]{\label{fig:top_4_8_optimal}
       	\includegraphics[width=0.46\columnwidth]{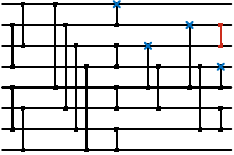}}\\
    \caption{Comparison of unary top-k selector derived from different unary sorters with 8 inputs.
    The top-k outputs are at the bottom.
    (a) and (b) are pruning bitonic sorters, while (c) and (d) are pruning optimal, thus smallest sorters~\cite{optimal_sorters}. 
    Each vertical segment represents one compare-and-swap unit in Fig.~\ref{fig:bitonic_sorting}.
    Black and red segments mark the mandatory and redundant compare-and-swap units in the unary sorter to implement unary top-k selector.
    Therefore, red units can be removed.
    Blue crosses mark these mandatory compare-and-swap units where only half of each unit is needed, i.e., one of the two outputs will not be used anymore.
    Blue crosses correspond to the dashed gates in Fig.~\ref{fig:neuron_uarch_catwalk}.
    x/y/z represent the number of total, mandatory, half compare-and-swap units.
    }
    \label{fig:top_k_n_selector}
\end{figure}

To implement unary top-k, we start with unary sorting in Fig.~\ref{fig:unary_sorting_tc}.
Unary sorting has all inputs sorted and is very straightforward to find the top-k of all inputs.
We show a few examples in Fig.~\ref{fig:top_k_n_selector}, assuming that the outputs are in an ascending order from top to bottom.
We evaluate two types of unary sorters: bitonic and optimal~\cite{optimal_sorters}. Bitonic sorters follow a structured bitonic pattern, while optimal sorters minimize the number of compare-and-swap units, achieving the lowest known count.
We use Algorithm~\ref{alg:top_k_pruning} to prune a given unary sorter and obtain the corresponding unary top-k selector.
\newcommand\mycommfont[1]{\textcolor{DarkGreen}{#1}}
\SetCommentSty{mycommfont}
\SetKwComment{Comment}{/* }{ */}
\begin{algorithm}
\caption{Top-k pruning.}\label{alg:top_k_pruning}
\KwIn{$\mathbfcal{S}$: a list of tuples for a unary sorter, with each tuple representing a compare-and-swap unit and ordered from left to right;
$n$: number of inputs; $k$: number of valid outputs.}
\KwOut{$\mathbfcal{T}$: a list of tuples for a unary top-k selector; $\mathbfcal{H}$: a list of tuples for the corresponding half compare-and-swap units.}
$\mathbfcal{M} = [n-k, ..., n-1]$\Comment*[r]{Initialize top $k$ outputs}
\For{(i, j) in reversed($\mathbfcal{S}$)}{
    \If{i or j in $\mathbfcal{M}$}{
        $\mathbfcal{T}$.insert($(i,j)$)\Comment*[r]{Insert tuple to front}
        $\mathbfcal{M}$.append($j$ or $i$)\Comment*[r]{Add the missing j or i}
    }
}
$\mathbfcal{L} = \mathbfcal{T} + [(n-k, n-k+1), ..., (n-2, n-1)]$ \Comment*[r]{Initialize tuple list}
\For{(i, j) in $\mathbfcal{L}$}{
    \If{i or j not in remainder\_node($\mathbfcal{L}$)}{
        $\mathbfcal{H}$.append($(i,j)$)\Comment*[r]{Add half unit}
    }
}
\Return $\mathbfcal{T}$, $\mathbfcal{H}$
\end{algorithm}

According to Fig.~\ref{fig:top_k_n_selector}, we have the following three observations.
First different unary sorters can yield identical top-k results with varying pruning efficiency; for top-2, bitonic and optimal sorters prune equally, but for top-4, bitonic prunes more.
Second, the final cost of unary top-k is independent of the cost reduction in compare-and-swap units.
For both top-2 and top-4, the final costs depend on both the original cost and the cost reduction.
Third, the higher the $k$, the higher the hardware cost.
From top-2 to top-4, less compare-and-swap units can be pruned, as well with the half compare-and-swap units (blue in Fig.~\ref{fig:top_k_n_selector}).

In general, we observe that optimal sorters yield better results, so we choose them for this work. However, a gap remains between unary sorting and unary top-k, as directly selecting the top k without full sorting could be even more resource-efficient.
This work focuses on optimal sorting-based solutions, leaving optimal top-k selection for future research.

\section{Experimental Setup}
\label{sec:Implementation}

We performed hardware evaluations using the NanGate45 standard cell library to obtain 45 nm CMOS implementation results and evaluation results. 
Design configurations of 16, 32, and 64 neuron inputs are chosen to enable comparisons across different scales. 
Synthesis carried out in Synopsys Design Compiler for three distinct design hierarchy configurations: (i) a stand-alone sorting/top-k stage, including unary bitonic sorters and optimal unary top-k, (ii) a sorting/top-k stage interfaced with a PC (a conventional design and a compact design), and (iii) bitonic sorting/optimal top-2 stage interfaced with a PC and augmented with a thresholding and firing unit, representing a SRM0-RNL neuron. 
At the final design hierarchy, all configurations are clocked at 400 MHz to ensure consistent timing assumptions. 

Designs are then placed and routed using Cadence Innovus with the NanGate45 cell library.
For these experiments, we again clock the designs at 400 MHz using a square floor plan with 70\% utilization for each input size to provide a consistent basis for comparison.

\section{Evaluation}
\label{sec:Evaluation}

This section evaluates the cost of our proposed \name neuron from both theoretical and experimental aspects.

\subsection{Gate Count Analysis}
Given the crux of using a lightweight unary top-k selector and PC to replace the original large PC, and the corresponding large design space, we seek to understand the potential of \name.
More specifically, we examine the gate count of multiple unary top-k and dendrite input designs.
We assume power-of-2 values for all $n$ (inputs) and $k$ (top-$k$ selections).

\begin{figure}[!t]
    \centering
    \subfloat[Gate count of unary top-k using Algorithm~\ref{alg:top_k_pruning}. 
    Light color at the bottom is for the number of effective gates that contribute to the functionality, and solid color at the top is for the removed gate in half compare-and swap units.
    When $n == k$, unary top-k becomes unary sorting with no pruning.
    ]{\label{fig:unary_top_k_cs_stacked_bargraph}
       	\includegraphics[width=\columnwidth]{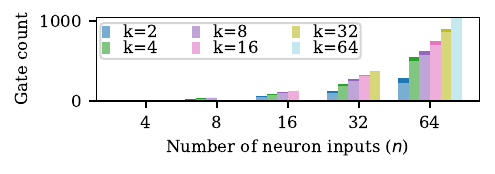}}\\
    \subfloat[Gate count of dendrite.
    The dendrite adopts unary top-k (using Algorithm~\ref{alg:top_k_pruning}) and compact PC ($n-1$ full adders for $n$ inputs).
    When $n == k$, the dendrite is just a large $n$-input compact PC without unary top-k.
    ]{\label{fig:unary_top_k_dendrite_gates_bargraph}
       	\includegraphics[width=\columnwidth]{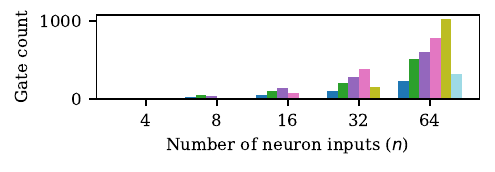}}\\
    \caption{Gate count analysis of unary top-k and dendrite. 
    }
    \label{fig:gate_count}
    \vspace{-5pt}
\end{figure}

Fig.~\ref{fig:unary_top_k_cs_stacked_bargraph} shows the gate count of different unary top-k designs.
We use optimal sorters to derive these unary top-k selectors.
The combined stacked bar is the total gate count for the remaining compare-and-swap units after pruning.

We observe that pruning compare-and-swap units significantly reduces hardware costs. Additionally, removing gates from half compare-and-swap units provides a smaller, but still helpful, optimization.
The gate saving trends for $n=\{16,32,64\}$
demonstrates the potential of unary top-k in reducing the dendrite cost, thus neuron cost, from a theoretical aspect.
Fig.~\ref{fig:unary_top_k_dendrite_gates_bargraph} shows the gate count for different dendrite designs.
We observe that when $k=2$, unary top-k offers gains in gate count, while larger $k$ values do not.
This is due to unary sorting-based unary top-k becomes more costly with larger $k$, as shown in Fig.~\ref{fig:top_k_n_selector}.

\subsection{Synthesis Results}

\subsubsection{Unary Top-K}
\begin{figure}[!t]
    \centering
    \subfloat[Area.]{\label{fig:unary_top_k_area_bargraph}
       	\includegraphics[width=\columnwidth]{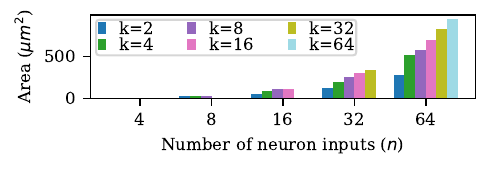}}\\
    \subfloat[Power.]{\label{fig:unary_top_k_power_bargraph}
       	\includegraphics[width=\columnwidth]{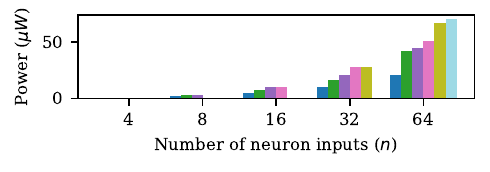}}\\
    \caption{Synthesis results of unary top-k.
    When $n == k$, unary top-k becomes unary sorting.
    }
    \label{fig:compare_unary_top_k}
    \vspace{-5pt}
\end{figure}

We synthesize unary sorting and unary top-k based on optimal sorting, which has the minimum number of compare-and-swap units known to date~\cite{optimal_sorters}.
Only $n=\{4, 8, 16, 32, 64\}$ are publicly available, and we leave the exploration of larger $n$ to future work.
The unary top-k selectors are obtained using the same method as in Fig.~\ref{fig:top_k_n_selector}.
We show the results in Fig.~\ref{fig:compare_unary_top_k}.
Fig.~\ref{fig:unary_top_k_area_bargraph} and Fig.~\ref{fig:unary_top_k_power_bargraph} show the synthesized area and power for different $n$ and $k$, and we observe graceful scaling when sweeping $n$ and $k$.

\subsubsection{Dendrite}
\begin{figure}[!t]
    \centering
    \subfloat[Area.]{\label{fig:dendrite_area_bargraph}
       	\includegraphics[width=\columnwidth]{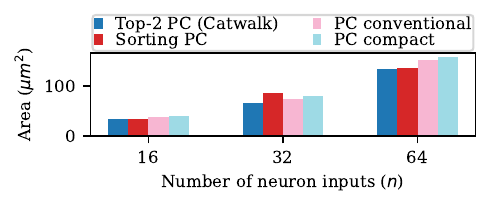}}\\
    \subfloat[Power.]{\label{fig:dendrite_power_bargraph}
       	\includegraphics[width=\columnwidth]{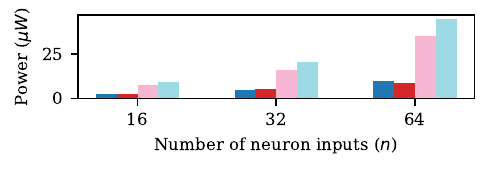}}\\
    \caption{Synthesis results of dendrite.
    Top-k here uses optimal sorters, while sorting use bitonic sorters.
    $n=\{16, 32, 64\}$ are studied, and $k$ is fixed to 2.
    }
    \label{fig:compare_dendrite}
    \vspace{-5pt}
\end{figure}

We also synthesize different dendrites, with results given in Fig.~\ref{fig:compare_dendrite}.
To avoid an exploding design space, we do not explore all dendrite combinations, but focus on the most efficient and available options.
As the sparsity in neurons can be as low as 0.1\%~\cite{silent_neuron, temporal_sparsity, measuring_sparsity}, we consider $k=2$ sufficient for the input count $n=\{16, 32, 64\}$.
We do not consider larger $n$, as no such optimal sorters are publicly available~\cite{optimal_sorters}.
Note that with $k=2$, the PC for top-k and sorting is just one full adder, as shown in Fig.~\ref{fig:neuron_uarch_catwalk}.

Three observations can be made.
First, unary top-k offers up to $1.17\times$ area savings over two PCs, aligning with theoretical gate count analysis.
However, unary sorting does not consistently show better or worse area.
Second, conventional PC (using an adder tree for accumulation) does not show worse area and power compared to compact PC (Fig.~\ref{fig:neuron_uarch_tnn}).
A conventional PC should have a larger cost in theory, but it is reasonable in the small scale that we are focusing on.
Third, both top-k and sorting show significant reduction in power consumption.
The leakage power of different designs remains similar, while top-k (and sorting) lowers the dynamic power significantly, boosting the power efficiency by up to $4.52\times$.

\subsubsection{Neuron}
\begin{figure}[!t]
    \centering
    \subfloat[Area.]{\label{fig:neuron_area_bargraph}
       	\includegraphics[width=\columnwidth]{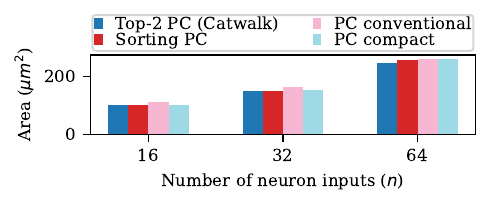}}\\
    \subfloat[Power.]{\label{fig:neuron_power_bargraph}
       	\includegraphics[width=\columnwidth]{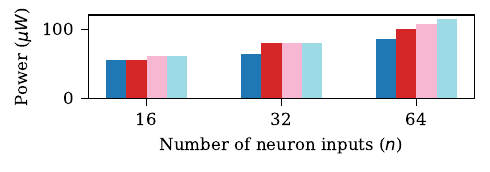}}\\
    \caption{Synthesis results of neuron.
    These neurons apply identical 5-bit accumulation and threshold implementation.
    Top-k here uses optimal sorters, while sorting use bitonic sorters.
    $n=\{16, 32, 64\}$ are studied, and $k$ is fixed to 2.
    Note that the PC compact neuron is from~\cite{tnn_microarch}, and the top-k PC neuron is our \name neuron.
    }
    \label{fig:compare_neuron}
    \vspace{-5pt}
\end{figure}

We further synthesize the full neuron, including dendrite, soma, and axon, with results given in Fig.~\ref{fig:compare_neuron}.
The experimental setup here is identical to that in dendrite evaluation.
\name (Top-2 PC) improves area and power by $1.05\times$ and $1.35\times$ over the neuron with a compact PC, and by $1.05\times$ and $1.17\times$ over the sorting-based neuron.
This aligns with insights from dendrite evaluation: while area reduction is limited, power improvement is significant.

\subsection{Place and Route Results}
In addition to synthesis, we also place and route all neurons, with
the results in Table~\ref{tab:compare_neuron_pnr}.
We observe that the leakage power of different design does not change too much, and \name's benefits mainly origins from the reduction in dynamic power.
Compared with the compact PC-based neuron (existing SRM0-RNL neuron design~\cite{tnn_microarch}), the area of \name is improved by $1.23\times$, $1.32\times$ and $1.39$; 
the power of \name is improved by $1.38\times$, $1.67\times$ and $1.86\times$ for $n=16,32,64$.
In general, more improvements are present with larger $n$.
Then the improvements are slightly more compared to the conventional PC-based neuron.
Also, within the evaluated range, \name indeed shows both area and power improvements compared to sorting PC-based neurons, indicating the importance of opting for top-k over sorting, despite identical functionality.

\begin{table}[!t]
\centering
\caption{Place-and-route results of different neurons in 45 nm CMOS. 
Neuron configurations are same as in Fig.~\ref{fig:compare_neuron}.}
\label{tab:compare_neuron_pnr}
\begin{tabular}{l|ccc|c}
\toprule
\multirow{2}{*}{\textbf{Neuron design}} & \multicolumn{3}{c|}{\textbf{Power ($\mu W$)}} & \textbf{Area} \\
\cmidrule(lr){2-4}
 & \textbf{Leakage} & \textbf{Dynamic} & \textbf{Total} & ($\mu m^2$) \\
\midrule
\multicolumn{5}{c}{$n=16, k=2$} \\
\midrule
PC conventional & 5.11 & 94.65 & 99.76 & 245.25 \\
PC compact~\cite{tnn_microarch} & 4.84 & 96.95 & 101.80 & 239.13 \\
Sorting PC & 4.28 & 70.11 & 74.39 & 197.64 \\
Top-k PC (\name) & 4.22 & 69.40 & 73.62 & 194.98 \\
\midrule
\multicolumn{5}{c}{$n=32, k=2$} \\
\midrule
PC conventional & 6.73 & 138.08 & 144.81 & 338.62 \\
PC compact~\cite{tnn_microarch} & 6.59 & 147.57 & 154.16 & 333.56 \\
Sorting PC & 5.73 & 88.24 & 93.97 & 256.42 \\
Top-k PC (\name) & 5.66 & 86.79 & 92.45 & 252.97 \\
\midrule
\multicolumn{5}{c}{$n=64, k=2$} \\
\midrule
PC conventional & 9.39 & 210.79 & 220.19 & 500.88 \\
PC compact~\cite{tnn_microarch} & 9.29 & 236.20 & 245.50 & 495.03 \\
Sorting PC & 8.12 & 129.59 & 137.71 & 364.15 \\
Top-k PC (\name) & 7.85 & 124.21 & 132.06 & 355.38 \\
\bottomrule
\end{tabular}
\end{table}

\section{Conclusion}
\label{sec:Conclusion}
This work identifies suboptimal efficiency of ramp-no-leak neurons in temporal neural networks, due to worse-case spike aggregation.
We propose to relocate the temporal spikes via unary top-k to reduce the hardware cost.
Unary top-k can be efficiently derived from unary sorting, with potential to reduce dendritic costs in neurons.
Through the use of lightweight unary top-k and parallel counter, we show that the post place-and-route area and power can be improved by up to $1.39\times$ and $1.86\times$, respectively, compared to existing neurons.

\bibliographystyle{plain}
\bibliography{ref}

\end{document}